\documentclass{ws-procs9x6}
\makeatletter
\DeclareRobustCommand*\cal{\@fontswitch\relax\mathcal}
\makeatother

\renewcommand{\vec}{\boldsymbol}
\newcommand{\be}{\begin{equation}}
\newcommand{\ee}{\end{equation}}
\newcommand{\bea}{\begin{eqnarray}}
\newcommand{\eea}{\end{eqnarray}}
\newcommand{\ba}{\begin{array}}
\newcommand{\ea}{\end{array}}

\newcommand{\Lb}{\left(}
\newcommand{\Rb}{\right)}

\newcommand{\eval}[1]{\left \langle #1 \right \rangle}

\def\eq#1{{Eq.~(\ref{#1})}}
\def\fig#1{{Fig.~\ref{#1}}}

\newcommand{\as}{\alpha_S}

\begin{document}

\title{COLOR CONFINEMENT\\ FROM FLUCTUATING TOPOLOGY}
\author{DMITRI E. KHARZEEV}

\address{Department of Physics and Astronomy, \\ Stony Brook University, New York 11794-3800, USA}
\address{Department of Physics and RIKEN-BNL Research Center, \\
Brookhaven National Laboratory, Upton, New York 11973-5000, USA}

\maketitle

\abstracts{
QCD possesses a compact gauge group, and this implies a non-trivial topological structure of the vacuum. 
In this contribution to the Gribov-85 Memorial volume, we first discuss the origin of Gribov copies and their interpretation in terms 
of fluctuating topology in the QCD vacuum. We then describe the recent work with {\mbox E. Levin} that links the confinement of gluons and color screening to the fluctuating topology, and 
discuss implications for spin physics, high energy scattering, and the physics of quark-gluon plasma.  
}


\section{The Gribov copies and knots in the QCD ether}\label{copies}

In 1977, Gribov pointed out \cite{GRCO} the fundamental ambiguity in non-Abelian gauge theory due to the existence of multiple solutions of the gauge fixing condition, known today as ``Gribov copies".  Much of the analysis was carried out in Coulomb gauge, but shortly afterwards Singer \cite{Singer:1978dk} showed that the problem persists in all other gauges, and thus no gauge fixing is possible in non-Abelian theories. Gribov has proposed to avoid the problem of copies by restricting the range of functional integration to small amplitude fields for which the solution to the gauge fixing condition is unique. He has demonstrated that such a restriction has a dramatic effect on the Coulomb interaction in QCD -- it becomes confining, with Coulomb potential rising linearly with the distance, see \cite{GRCREV,Dokshitzer:2004ie} for reviews.

\vskip0.3cm  
In this contribution (based on a recent work with E. Levin \cite{Kharzeev:2015xsa}), we will take a different route -- namely, we will argue that Gribov copies naturally arise from the compactness of the gauge group, and instead of eliminating them, we have to build a picture of the vacuum that accommodates their existence. The physical quantities in non-Abelian theories should thus be evaluated by summing over Gribov copies. When applied to the gluon propagator in QCD, this recipe leads to the confinement of gluons at large distances \cite{Kharzeev:2015xsa}. 
\vskip0.3cm

In fact, shortly after Gribov pointed out the existence of gauge copies, Jackiw, Muzinich and Rebbi \cite{Jackiw:1977ng} showed that the copies, and associated with them discontinuities of gauge potentials are necessary to accommodate the vacuum configurations with arbitrary Chern-Pontryagin index. In an even number of space-time dimensions, such as our $(3+1)$ dimensional Minkowski space, the Chern-Pontryagin index is associated with topology-changing transitions. The discussion of topology of gauge theory is usually restricted to semi-classical topological solutions (instantons, monopoles, dyons, ...) and is thus in the domain of ``non-perturbative QCD" traditionally ignored by the practitioners of perturbation theory. However a meaningful perturbartive expansion is based on fixing the gauge, and the existence of Gribov copies thus implies the impossibility to isolate perturbation theory from the topological nature of QCD. 
\vskip0.3cm

The relevance of topology in gauge theories stems from the compactness of the gauge group. Non-Abelian gauge theories always possess a compact gauge group, but even in the Abelian case we may choose to treat the $U(1)$ group as compact -- and in this case the theory acquires a non-trivial topological structure, and the confined phase, see e.g.\cite{Panero:2005iu}. This fact suggests a link between topology and confinement. Compact gauge groups in general possess a homotopy map to the space-time manifold. For example, the map from the ${\rm SU(2)} \subset {\rm SU(N)}$ subgroup of the gauge group to the Euclidean space-time sphere ${\rm S^3}$ describes the classical instanton solution \cite{Belavin:1975fg}. 
In Minkowski space-time, instanton solutions represent the tunneling events connecting the degenerate vacuum states with different Chern-Simons numbers 
\be\label{CS_num}
X (t) = \int d^3 x\ K_0 (x, t),
\ee
where $K_0$ is the temporal component of topological current 
\be\label{CS_cur}
K_\mu = \frac{1}{16 \pi^2} \epsilon_{\mu\nu\rho\sigma} A^{\nu,a} \left(\partial^\rho A^{\sigma,a} + \frac{1}{3}g C^{abc} A^{\rho}_b A^{\sigma}_c \right) ;
\ee 
we have absorbed the coupling constant in the definition of the field.
\vskip0.3cm

The density of Chern-Pontryagin number mentioned above is the divergence of topological current (\ref{CS_cur})
\be\label{CP}
Q(x) = \partial^\mu K_\mu ,
\ee
and is thus given by
\be
Q(x) = \frac{1}{32 \pi^2} F_{\mu\nu}(x) \tilde{F}^{\mu\nu}(x) .
\ee
The Chern-Pontryagin index is $\int d^4x\ Q(x) \equiv \nu$; for finite action field configurations $\nu$ is an integer. 
As shown in Refs \cite{Jackiw:1977ng,Singer:1978dk}, the presence of Gribov copies is necessary to accommodate configurations with an arbitrary (integer) value of Chern-Pontryagin index.
\vskip0.3cm

Prior to proceeding further, let us develop an intuitive picture of the vacuum structure implied by the topology of the compact gauge group. For simplicity, let us begin with the case of Abelian gauge theory -- a familiar electrodynamics. In this case we limit ourselves to the first term in (\ref{CS_cur}). The corresponding Abelian Chern-Simons number (\ref{CS_num}) 
has been known since Gauss's work in the XIX century -- it is a so-called ``magnetic helicity" introduced in magnetohydrodynamics by Woltjer\cite{Woltjer} and Moffatt\cite{Moffatt}: 
\be
{\cal H}(t) = \int d^3 x\ \vec{A} \cdot \vec{B} ;
\ee
to be precise, the Chern-Simons number $X(t)$ (\ref{CS_num}) is equal to ${\cal H}(t)$ divided by $16 \pi^2$. 
\vskip0.3cm

Magnetic helicity measures the ``knottedness" of the lines of magnetic flux. To quantify this statement and to characterize the topology of magnetic flux, let us split magnetic field into $N$ magnetic flux tubes with magnetic fluxes $\phi_i$. The magnetic helicity can then be written down \cite{Moffatt,BF84,MR92,Berger} in terms of the geometric invariant of the links between these flux tubes:
\be\label{mag_top}
{\cal H}(t) = \sum_{i=1}^N \phi_i^2\ {\cal S}_i + 2 \sum_{i,j} \phi_i \phi_j\ {\cal L}_{ij}, 
\ee
where ${\cal S}_i$ is the C\u{a}lug\u{a}reanu-White self-linking number, and ${\cal L}_{ij}$ is the Gauss linking number\footnote{The linking numbers in (\ref{mag_top}) do not always detect the topology of the link; the development of the appropriate universal knot invariants is a very active area in modern mathematics.}. 
\vskip0.3cm

In infinite space, the solution of the Maxwell equations characterized by a non-zero value of (\ref{mag_top}) is very simple -- it is a circularly polarized plane wave. The magnetic helicity is a parity--odd quantity, and the parity transformation turns a left-handed circularly polarized wave into the right-handed one. In the presence of the boundaries, the solution\cite{Hopf} becomes a Hopfion with linked lines of fields characterized by non-zero value of the second term on the r.h.s. of (\ref{mag_top}). 
\vskip0.3cm

The notion of magnetic helicity is extensively used in magnetohydrodynamics, where it is conserved in the absence of dissipation -- in geometrical picture implied by (\ref{mag_top}), this means that the lines of magnetic field are ``frozen" into the fluid and do not reconnect. Because the time derivative of magnetic helicity is
\be
\frac{d}{d t} {\cal H}(t) = \frac{d}{d t} \int d^3 x\ \vec{A} \cdot \vec{B} = - 2 \int d^3 x\ \vec{E} \cdot \vec{B}
\ee
from the definition of electric field $\vec{E}$, we see that conservation of magnetic helicity implies $\int d^3 x\ \vec{E} \cdot \vec{B} = 0$. In electrodynamics, this quantity is nothing but the Chern-Pontryagin index $\nu$ defined above, since  $F_{\mu\nu} \tilde{F}^{\mu\nu} = - 4 \vec{E} \cdot \vec{B}$. The gauge field configurations with non-zero Chern-Pontryagin numbers (such as the ones signaled by the presence of Gribov copies in non-Abelian theory) thus have an important role: they induce {\it reconnections of magnetic flux}.  
\vskip0.3cm

In classical magnetohydrodynamics, reconnections of magnetic flux are often dramatic events -- for example, in the corona of the Sun they are responsible for solar flares and coronal mass ejections spanning millions of miles (for review, see\cite{eject}). In quantum theory, the physics during a magnetic reconnection can be described by adding the Chern-Simons term $\mu_5 K^0$ to the action of electrodynamics\cite{Kharzeev:2009fn}, where $\mu_5$ is the chiral chemical potential -- since magnetic helicity $K^0$ is parity-odd, $\mu_5$ is parity-odd as well. The equations of motion of the resulting Maxwell-Chern-Simons theory then exhibit the chiral magnetic effect \cite{Fukushima:2008xe} (see \cite{Kharzeev:2013ffa} for review and references) -- the electric current $\vec{J} = e^2/2\pi^2\ \mu_5 \vec{B}$. The spectrum of photons in this theory differs from the one in usual Maxwell electrodynamics and exhibits a ``chiral magnetic instability" -- the photon energy acquires an imaginary part at soft momenta $k \leq (e^2/2\pi) \mu_5$ reflecting a growth or decay of the photon field\cite{RW,Tso,JS,Akamatsu:2013pjd}. 
\vskip0.3cm

At high momenta $k \gg (e^2/2\pi) \mu_5$ the photons behave as a free quanta of perturbation theory, but in the infrared region the behavior is totally different -- the photons either cease to propagate (are confined), or are produced (this is a quantum analog of the solar flare induced by a magnetic reconnection). In the case of the vacuum, which is by definition the state with the lowest energy, the radiation of photons is clearly impossible and the physical choice for the imaginary part of the photon energy corresponds to the photon decay. If the magnetic reconnections take place in the vacuum with a certain rate per unit time per unit volume, then this rate will determine the characteristic length over which the soft photon is allowed to propagate. We will see in the next section that this may be exactly what happens in QCD to soft gluons. 
\vskip0.3cm

The emerging picture of the vacuum filled by the knots is remarkably close to the one proposed by Lord Kelvin in 1867\cite{Kelvin}. Inspired by Helmholtz's work on vortex motion in perfect liquid, Kelvin proposed that the observed atoms were in fact the knots of vortices in the ether. He wrote\cite{Kelvin}:  {\it ...the endless variety of [knotted or knitted vortex atoms] is infinitely more than sufficient to explain the varieties and allotropies of known simple bodies and their mutual affinities.} While we now know that atoms are definitely not knots in the ether, Kelvin's prophecy may still apply to the structure of the QCD vacuum.

\section{Veneziano's ghost and fluctuating topology in the QCD vacuum}\label{aba:sec1}

The relevance of non-trivial topology in the QCD vacuum is revealed in the physical spectrum of the theory through the chiral anomaly\cite{Adler:1969gk,Bell:1969ts} and the resulting non-conservation of the axial current:
\be\label{chan}
\partial_\mu J_A^\mu = 2 N_f Q(x) + \sum_f (2 i m_f) \bar{q}_f \gamma_5 q_f,
\ee
where $m_f$ are the masses of quarks, $N_f$ is the number of flavors, and $Q(x)$ is the density of Chern-Pontryagin number given by  (\ref{CP}). The vacuum fluctuations in topology and the chiral anomaly thus lead to the explicit breaking of $U_A(1)$ symmetry, and are responsible\cite{VEN,Witten:1979vv} for the anomalously large mass of the would-be Goldstone boson $\eta'$, as we discuss below.
\vskip0.3cm

Veneziano \cite{VEN} has demonstrated that the fluctuations of topology in the vacuum can be captured by introducing a massless ``ghost" in the correlation function of the gauge-dependent topological current (\ref{CS_cur}):
\be \label{VEN}
K_{\mu \nu}\Lb q \Rb\,\equiv \,i \,\int d^4 x\ e^{ i q x}\,\eval{ T\{K_\mu\Lb x\Rb K_\nu \Lb 0\Rb\}}\,\,\xrightarrow{ q^2 \,\ll \,\mu^2}\,\,- \frac{\mu^4}{q^2}\ g_{\mu \nu},
\ee
where $\mu^4 \equiv \chi_{top}$ is the topological susceptibility of pure Yang-Mills theory. Note that the r.h.s. of \eq{VEN} has the ``wrong" sign, which means that the ghost does not describe a propagating degree of freedom. Therefore the ghost cannot be produced in a physical process; however the couplings of the ghost (that describe the effect of topological fluctuations) can affect physical amplitudes. A similar ``dipole" ghost had been earlier introduced by Kogut and Susskind \cite{Kogut:1974kt} in the analysis of axial anomaly in the Schwinger model. This procedure has been demonstrated to solve the $U_A(1)$ problem in QCD \cite{Weinberg:1975ui,Witten:1979vv}.
\vskip0.3cm

The physical meaning of \eq{VEN} becomes apparent if one compares it to the correlation function of the electron's coordinate $x(t)$ in a crystal, following Dyakonov and Eides \cite{Dyakonov_Eides}:
\be\label{cryst}
i \int dt \ e^{i \omega t}\ \langle T\{ x(t) x(0) \}\rangle \  \xrightarrow{\omega \to 0} - \frac{1}{\omega^2\ m^*} = - \frac{1}{\omega^2}\  \frac{\partial^2 E(k)}{\partial k^2} \biggr\rvert_{k = 0} ,
\ee
where $E = k^2/ 2 m^*$ is the energy of an electron with an effective mass $m^*$ and quasi-momentum $k$ in a crystal. The emergence of the pole in \eq{cryst} signals the possibility of electron's propagation in the periodic potential of the crystal by means of  tunneling. It is important to note that the pole emerges not just from a single tunneling event (corresponding to the isolated instanton in QCD), but sums up the effect of many tunnelings throughout the crystalline lattice. 
\vskip0.3cm
The analogy between \eq{VEN} and \eq{cryst} becomes even more apparent if we choose the frame with $q^\mu = (\omega, 0)$ and use the analog of coordinate given by \eq{CS_num}.  
The expression \eq{VEN} then takes the form that is completely analogous to \eq{cryst}:
\be\label{ghost}
i \int dt\ e^{ i \omega t}\,\langle T\{X(t) X(0)\} \rangle \,\,\xrightarrow{\omega \to 0} - \frac{\mu^4}{\omega^2}\ V = - \frac{1}{\omega^2}\ \frac{\partial^2 E(\theta)}{\partial \theta^2} \biggr\rvert_{\theta = 0} ,
\ee
where $V$ is the volume of the system, and $E(\theta) = \epsilon(\theta) V$ is the energy of the Yang-Mills vacuum. 
The energy density of the vacuum $\epsilon(\theta)$ is a periodic function of the $\theta$ angle that is analogous to the quasi-momentum $k$ in \eq{cryst}. At small $\theta$, we can expand $\epsilon(\theta)$:
\be
\epsilon(\theta) \simeq \mu^4\ \frac{\theta^2}{2} ,
\ee 
which exhibits the physical meaning of the quantity $\mu^4$ on the r.h.s. of (\ref{ghost} -- it is the topological susceptibility $\chi_{top} = \mu^4$ of the Yang-Mills theory. Note that a term linear in $\theta$ is forbidden by P and CP invariances of QCD.
\vskip0.3cm
The expression (\ref{ghost}) clearly exhibits the role of the ghost -- it describes the effect of topology-changing transitions characterized by non-vanishing Chern-Pontryagin indices in the vacuum. In the Abelian case presented in the previous section, the ghost would describe magnetic reconnections occurring with the rate (per unit volume, unit time) proportional to $\mu^4$.
\vskip0.3cm
This interpretation becomes even more apparent if we consider the correlation function of the Chern-Pontryagin number density $Q(x)$ (\ref{CP}) implied by (\ref{VEN}), as we will now explain.
The assumption of the ghost existence (\ref{VEN}) implies the following Minkowski-space correlation function for $Q(x) = \partial^\mu K_\mu$:
\be\label{wn}
i \,\lim_{q \to 0} \int d^4 x\ e^{ i q x}\,\eval{ T\{Q\Lb x\Rb Q\Lb 0\Rb\}}\, = - \mu^4 < 0 .
\ee
The quantity $Q(x)$ is gauge-invariant, and excites from the vacuum physical states, the lightest of which is the flavor-singlet $\eta'$ meson\footnote{For simplicity we do not discuss here the $\eta' - \eta$ mixing.}. By writing down the spectral representation for the correlator (\ref{wn}) it is easy to see that the contribution of these physical states to the r.h.s. should be positive:
\be\label{wn-pos}
i \,\lim_{q \to 0} \int d^4 x\ e^{ i q x}\,\eval{ T\{Q\Lb x\Rb Q\Lb 0\Rb\}}\, = \int \frac{dM^2}{2 \pi}\ \frac{\rho(M^2)}{M^2} > 0.
\ee
\vskip0.3cm

How can we reconcile (\ref{wn}) and (\ref{wn-pos})? The key to the answer stems from the observation that in the physical world (in which the dispersion integral (\ref{wn-pos}) is written), the theory possesses light quarks of mass $m_f$, and in the chiral limit of massless quarks the correlation function of topological charge in (\ref{wn-pos}) must vanish. This is because at $m_f =0$ the topological charge is a full divergence of the gauge-invariant axial current (see (\ref{chan})). 
\vskip0.3cm

For non-flavor-singlet axial currents $J_{A, NS}^\mu$ that do not mix with gluon states, the vanishing of the correlation function of $\partial_\mu J_{A, NS}^{\mu}$ at zero momentum is assured by the Goldstone theorem, since  the matrix elements of the divergence of axial current between the vacuum and  the Goldstone boson (e.g. a pion) 
\be
\langle 0| \partial_\mu J_{A, NS}^{\mu} | \pi \rangle = m_{\pi}^2 f_{\pi}
\ee
vanish in the chiral limit of massless quarks as $m_\pi \to 0$. 
\vskip0.3cm
In the case of the flavor singlet axial current producing from the vacuum the $\eta'$ meson with matrix elements $
\langle 0| J_{A}^{\mu} | \eta' \rangle = \sqrt{2 N_f}\ p^\mu f_{\eta'}
$
and
$
\langle 0| \partial_\mu J_{A, NS}^{\mu} | \pi \rangle = \sqrt{2 N_f}\ m_{\eta}^2 f_{\pi} 
$
(normalization is chosen in accord with (\ref{chan})), 
the situation is different because of the presence of topological fluctuations expressed by the relation (\ref{wn}). To construct the physical correlation function in QCD, we thus have to add the r.h.s. of (\ref{wn}) and (\ref{wn-pos}). Saturating the dispersion integral by the $\eta'$ peak and choosing in accord with (\ref{chan}) 
$
\langle 0| Q | \eta' \rangle = 1/\sqrt{2 N_f}\ m_{\eta}^2 f_{\pi}
$
and
$\rho(M^2) = 4 N_f\ m_{\eta'}^4 f_{\eta'}^2\ 2\pi \delta(M^2 - m_{\eta'}^2)$, we get
\be
i \,\lim_{q \to 0} \int d^4 x\ e^{ i q x}\,\eval{ T\{Q\Lb x\Rb Q\Lb 0\Rb\}}_{\rm QCD} = - \mu^4 + \frac{m_{\eta'}^2 f_{\eta'}^2}{4 N_f}  = 0,
\ee
from which we immediately get the Witten-Veneziano relation\cite{VEN,Witten:1979vv} for the $\eta'$ mass:
\be
m_{\eta'}^2 f_{\eta'}^2 = 4 N_f\ \mu^4 = 4 N_f\ \frac{\partial^2 \epsilon(\theta)}{\partial \theta^2} \biggr\rvert_{\theta = 0}^{YM} ;
\ee
we have indicated explicitly that the topological susceptibility has to be computed for Yang-Mills theory (without light quarks). 
\vskip0.3cm

Subtracting a constant from the r.h.s. of (\ref{wn-pos}) may seem strange -- however, we have to keep in mind that
 the dispersion integral in (\ref{wn-pos}) diverges in the high mass region that we have omitted, since on dimensional grounds $\rho(M^2) \sim M^4$ (as reproduced by the gluon loop) -- therefore, it requires subtractions, and this is what the $\mu^4$ term represents\footnote{The T-ordered product in (\ref{wn-pos}) has to be understood as a Wick, and not Dyson, one\cite{Dyakonov_Eides}.}. The Witten-Veneziano prescription described above thus implies that the correlation function (\ref{wn}) should contain a contact negative term, 
$\eval{ T\{Q\Lb x\Rb Q\Lb 0\Rb\}} \sim - \mu^4\ \delta^4(x)$.  There is an ample evidence for this contact term from the lattice simulations, see Fig.1 from \cite{Horvath} (the correlator shown in Fig.1 is Euclidean, hence the sign difference). 
\vskip0.3cm

The physical interpretation of the contact term is suggested by our discussion in section \ref{copies} -- it describes the ``white noise" of topological reconnections, with strength given by the topological susceptibility $\mu^4$. Just like the introduction of white noise in kinetic theory allows one to describe the effect of the medium on a test particle, the ghost allows to capture the impact  of fluctuating topology. Compared to ``microscopic" approaches to QCD vacuum, such as the instanton liquid model and its descendants (for review, see \cite{Schafer:1996wv}), 
the approach described here does not identify the dynamics of topological transitions -- instead, we parameterize the rate of transitions by a single dimensionful parameter, topological susceptibility. On the other hand, it does not rely on a (questionable) semiclassical approximation. 
\vskip0.3cm

Usually, the discussion of topological effects is limited to the $U_A(1)$ problem and correlation functions of color-singlet correlators, as presented above. However, the fluctuating topology should also affect the propagation of gluons, similarly to the case of photons discussed in the previous section. We will now see that this is indeed the case. The Veneziano ghost defined by (\ref{VEN}) will allow us to address this problem in QCD without resorting to the complicated microscopic dynamics of topological transitions.   

\vskip0.3cm

\begin{figure}
\begin{center}
\includegraphics[width=10cm]{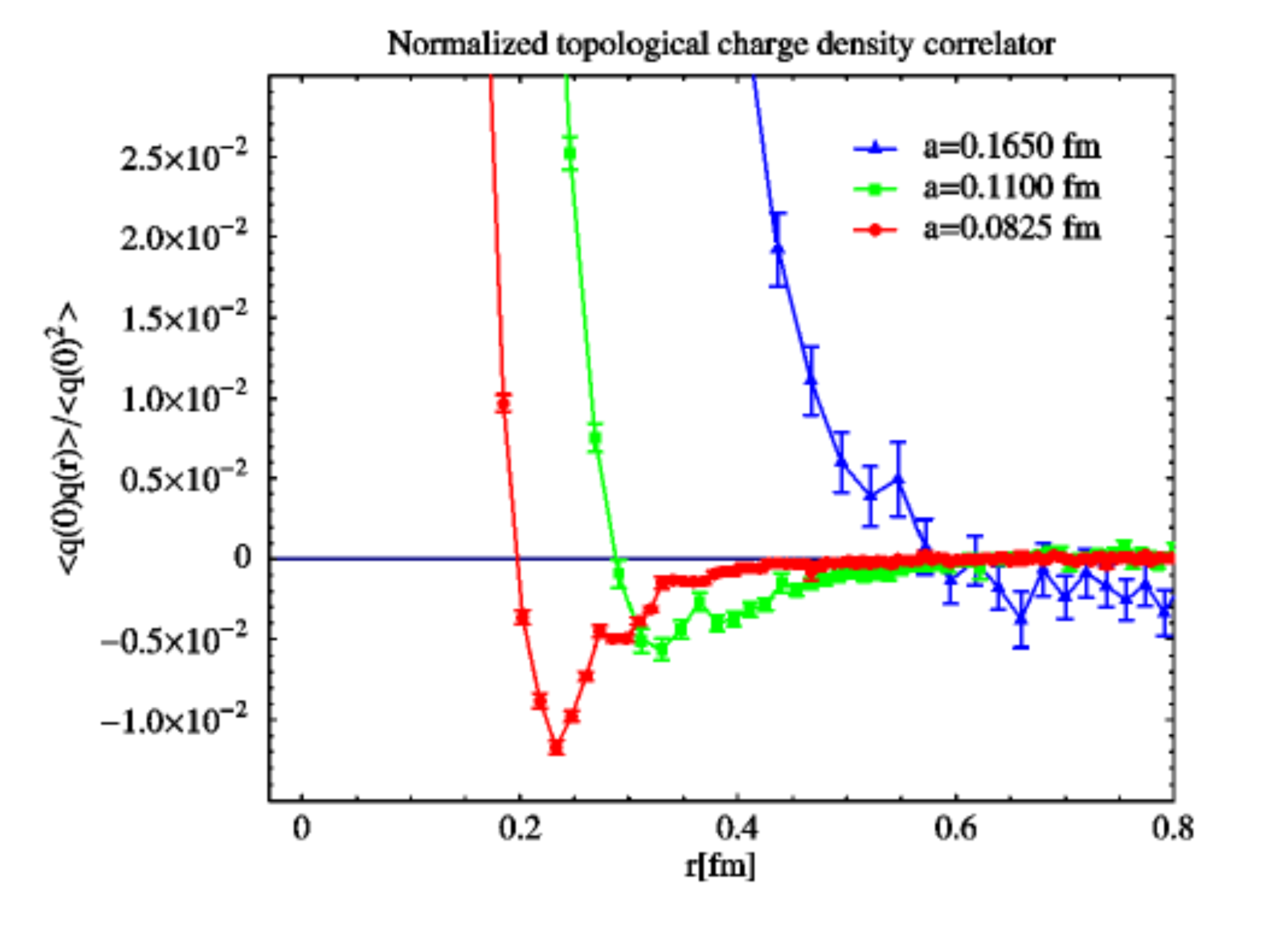}
\end{center}
\caption{The lattice QCD data on the correlation function of Chern-Pontryagin charge density 
at three lattice spacings $a$; from$^{31}$.}
\label{lat-top}
\end{figure}

The observation made in our recent paper\cite{Kharzeev:2015xsa} with E. Levin is that (\ref{VEN}) and (\ref{CS_cur}) define an effective ghost-gluon-gluon vertex  $\Gamma_\mu\Lb q,p\Rb$. Since the topological current (\ref{CS_cur}) is a pseudo-vector, the ghost is a pseudo-vector ``particle" -- so the interaction of a gluon with, say, a left circular polarization with the ghost will transform it into a right circularly polarized gluon. Because the QCD vacuum is P and CP even, the ghost can only appear in the loops. 
\vskip0.3cm

\begin{figure}
\begin{center}
\includegraphics[width=10cm]{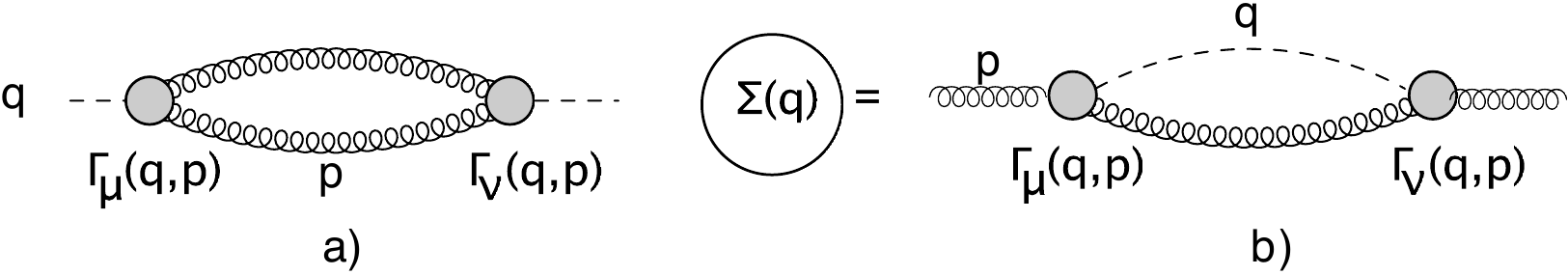}
\end{center}
\caption{\fig{kk}-a : \eq{VEN} in momentum representation; helix lines represent gluons and the dashed line depicts the ghost. \fig{kk}-b : the gluon dressed by the interactions with the ghost: a ``glost", see \eq{SIGMA} for the corresponding self-energy expression.}
\label{kk}
\end{figure}

Using this vertex, we can re-write the correlator  \eq{VEN} at small $q^2$ as follows (see \fig{kk}-a): 
\be \label{KMOM}
K_{\mu \nu}\Lb q \Rb\,\,=\,\,\frac{1}{( 2 \pi)^4 i}\int d^4 p ~\Gamma_\mu\Lb q,p\Rb\frac{1}{ p^2 (q -p )^2}\, \Gamma_\nu\Lb q,p\Rb\,=\,\,- \frac{\mu^4}{q^2}\,g_{\mu \nu} ,
\ee
from which we  find that
\be \label{GG}
 \Gamma_\mu\Lb q,p \Rb\,\Gamma_\nu\Lb q,p\Rb\,\,\propto\, - \frac{\mu^4}{p^2}\,g_{\mu \nu}\ ,\,\mbox{for}\,\,q \,\leq p .
 \ee 
The vertices  $\Gamma_\mu\Lb q,p\Rb$ describe the interactions of soft gluons with topological fluctuations. 
\vskip0.3cm

Indeed, the gluon  propagator is now the solution of the Dyson-Schwinger equation with $\Sigma\Lb p\Rb$
given by 

\be \label{SIGMA}
\Sigma_{\mu\nu}\Lb p\Rb~=~\frac{1}{( 2 \pi)^4 i}\int d^4 q ~\Gamma_\mu\Lb q,p\Rb\frac{1}{ q^2 (q -p )^2}\, \Gamma_\nu\Lb q,p\Rb
\ee
where $1/q^2$ is the propagator of the ghost. In evaluating the integral of \eq{SIGMA} we assume that $q \ll\,p$ since 
$\Gamma_\mu$'s describe the non-perturbative effects at small momenta. Therefore,
\be \label{SIGMA1}
\Sigma_{\mu\nu}\Lb p\Rb~\simeq~\frac{4\pi }{( 2 \pi)^4 }\int^p_0 d q^2 q^2 ~\Gamma_\mu\Lb q,p\Rb\frac{1}{ q^2 p^2}\, \Gamma_\nu\Lb q,p\Rb\,\,\,=\,\,- g_{\mu \nu} \frac{\mu^4}{p^2} ,
\ee
where we used \eq{GG}.
\vskip0.3cm

We can now write down the Schwinger-Dyson equation for the gluon propagator \footnote{This is the expression in Feynman gauge; since the self-energy $\Sigma_{\mu\nu}$ does not change the polarization of the gluon, the Dyson-Schwinger series can be re-summed in the usual way.} $G_{\mu \nu}\Lb p\Rb\,\,=\,\,g_{\mu \nu}\,G\Lb p\Rb$ : 
\be \label{EQ}
G\Lb p \Rb\,\,=\,\,\frac{1}{p^2} \, + \,\frac{1}{p^2} \Sigma\Lb p\Rb \,G\Lb p \Rb
\ee
with the solution
\be \label{SOL1}
G\Lb p \Rb\,\,=\,\,\frac{1}{p^2 - \Sigma\Lb p\Rb}\,\,=\,\,\frac{1}{ p^2 + \frac{\mu^4}{p^2}}
\ee
\vskip0.3cm

The propagator (\ref{SOL1}) has some remarkable properties. First, 
   $G\Lb p \Rb$ has no gluon pole in the physical region. Instead, this propagator has only complex poles at $p^2 \,=\,\pm i \mu^2$.  As a result, gluons cannot propagate at large distances  -- in other words, they are confined. 
   Second, the propagator of the type of \eq{SOL1} as shown by Gribov \cite{GRCO} and Zwanziger \cite{Zwanziger:1989mf} eliminates the gauge copies. Note however that 
   we have obtained this propagator not by restricting the range of functional integration to the region where the copies do not appear, but by taking account of all copies into account through the topological fluctuations.  
   \vskip0.3cm
   
   Let us illustrate the meaning of our result by appealing again to the analogy with the crystal, see (\ref{cryst}). The Gribov copies in this example are analogous to the electron states localized around different sites in the crystalline lattice. When we discuss the wave function of the electron around one of the sites, in general we cannot ignore the presence of the tunneling throughout the lattice. The tunneling can be ignored only if we limit ourselves to the tightly bound electron levels. However the tunneling can be consistently taken into account if instead of an individual bound state we consider the electron bands. Electron energies in these bands acquire a finite width, similarly to what we have observed for the gluons. 
\vskip0.3cm   
   
Our propagator \eq{SOL1} results from the admixture of the ghost to the perturbative gluon (see \fig{kk}-b), with an amplitude defined by the Yang-Mills topological susceptibility $\mu^4$.  In paper\cite{Kharzeev:2015xsa}, we proposed to call this  coherent mixture of a {\bf gl}uon and a gh{\bf ost}  a ``{\bf glost}".  Unlike the ghost, the ``glost" can be produced in a physical process, but unlike the perturbative gluon, it is confined and cannot propagate at distances larger than $\sim \mu^{-1} \sim 1$ fm.

\begin{figure}
\includegraphics[width=8cm]{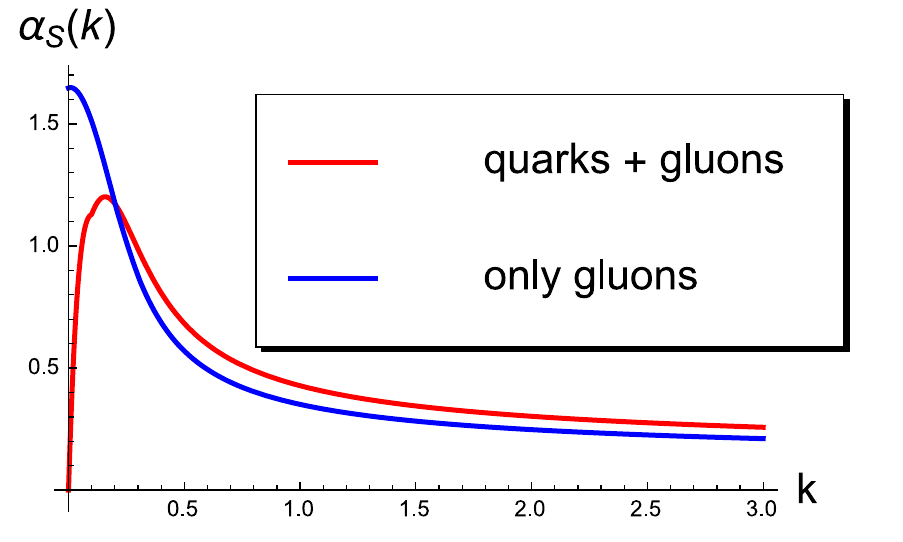}
\caption{The  running constant $\as$ as a function of momentum $k$ in our model for QCD with light quarks (red line that goes to zero at $k\to0$) and gluodynamics (blue line); from$^5$. }
\label{as}
\end{figure}

\vskip0.3cm
 At distances $\sim \mu^{-1} \sim 1$ fm, the use of the glost propagator also gives rise to the screening of color charge\cite{Kharzeev:2015xsa}, leading to the freezing of the effective coupling in the infrared ${\rm IR}$ limit for pure gauge theory, or to the  vanishing of the coupling in the ${\rm IR}$ in QCD with light quarks, see Fig. \ref{as}. It has been shown by Dokshitzer \cite{DOK} that the experimental data indicate the relative smallness of the QCD coupling in the IR region:  
     \be \label{QCD7}
    \alpha_0\,\,=\,\,\frac{1}{\mu_I}\,\int^{\mu_I} d k \,\as\Lb k \Rb\,\,\approx\,\,0.5 \,\,\,\,\,\mbox{for}\,\,\,\mu_I\,\,=\,\,2\,\,{\rm GeV} .
    \ee
 In  our approach we get $\alpha_0 \,\simeq\,0.59$ for renormalization point $k = M_Z$, in reasonable agreement with (\ref{QCD7}). 
  
 Note that close to the deconfinement transition, the topological susceptibility vanishes reflecting the restoration of $U_A(1)$ symmetry \cite{Bonati:2013tt,Kharzeev:1998kz}. Since at $\mu \to 0$ the gluon propagator becomes perturbative, the restoration of $U_A(1)$ symmetry and deconfinement should occur at the same temperature in agreement with the lattice data \cite{Bonati:2013tt}; however close to $T_c$ the non-perturbative interactions induced by topological fluctuations are important. It would be interesting to explore the implications of this approach to QCD thermodynamics, and to the rates of electromagnetic emission from the quark-gluon plasma -- this work has already begun \cite{Bandyopadhyay:2015wua}.

\vskip0.3cm
It is of fundamental interest to establish the microscopic dynamics responsible for the fluctuating topology captured by the ghost. A recent study within the ``deformed QCD" attributes these fluctuations to the topological order in the vacuum \cite{Zhitnitsky:2013hs}. 
The imaginary part of the gluon self-energy signals an instability w.r.t. decay to the true ground state -- what is the structure of this ground state? The clue to the answer may be provided by the Abelian example considered in section 1 -- there, the true ground state of the photon field is the (self-)linked knot of magnetic flux. The knots repeatedly emerge in studies of the ground state of non-Abelian gauge theories\cite{Faddeev:2008qe,Faddeev:2001dda,Gorsky:2015toa}, but it is clear that much remains to be done. 
\vskip0.3cm

In the meantime, the simplistic approach described above captures the link between the confinement and the fluctuating topology in the QCD vacuum, and provides a practical way of computing power-suppressed corrections to QCD amplitudes. In particular, the glost propagator leads to the exponential fall-off of the high-energy hadron scattering amplitude at large impact parameters needed to satisfy the Froissart bound; this can solve the long-standing problem of the perturbative approach in describing high energy scattering \cite{Kovner:2001bh}. In QCD amplitudes  the coupling to the ghost flips the gluon and (through loops) quark polarizations, and can give rise to spin asymmetries \cite{Kang:2010qx} that are different from the usual perturbative approach.  

\section{Acknowledgments}
I am indebted to Yuri Dokshitzer and Julia Nyiri for the invitation to contribute to the Gribov-85 Memorial volume. The work presented here was done together with Eugene Levin; I am grateful to him for the enjoyable collaboration. This work was supported in part by the U.S. Department of Energy under Contracts
DE-FG-88ER40388 and DE-SC0012704. 



\end{document}